\documentclass[12pt]{article}
\pdfoutput=1
\usepackage{graphics, color}
\usepackage{graphicx}
\usepackage{amsmath}
\usepackage{amssymb}
\usepackage{amsfonts}

\newcommand{\sect}[1]{\section{#1}\setcounter{equation}{0}}

\def\gsim{\, \rlap{$>$}{\lower 1.1ex\hbox{$\sim$}}\,}
\def\lsim{\, \rlap{$<$}{\lower 1.1ex\hbox{$\sim$}}\,}

\begin{document}


\begin{titlepage}
\bigskip
\bigskip\bigskip\bigskip
\centerline{\Large Black Holes: Complementarity or Firewalls?}
\bigskip\bigskip\bigskip
\bigskip\bigskip\bigskip

 \centerline{{\bf Ahmed Almheiri,}\footnote{\tt ahmed@physics.ucsb.edu}*
 {\bf Donald Marolf,}\footnote{\tt marolf@physics.ucsb.edu}*${}^\dagger$
 {\bf Joseph Polchinski,}\footnote{\tt joep@kitp.ucsb.edu}${}^\dagger$
 and {\bf James Sully}\footnote{\tt sully@physics.ucsb.edu}*}
\medskip
\centerline{\em *Department of Physics}
\centerline{\em University of California}
\centerline{\em Santa Barbara, CA 93106}
\medskip
\centerline{\em ${}^\dagger$Kavli Institute for Theoretical Physics}
\centerline{\em University of California}
\centerline{\em Santa Barbara, CA 93106-4030}\bigskip
\bigskip\bigskip


\begin{abstract}
We argue that the following three statements cannot all be true: (i)  Hawking radiation is in a pure state, (ii)  the information carried by the radiation is emitted from the region near the horizon, with low energy effective field theory valid beyond some microscopic distance from the horizon, and (iii)  the infalling observer encounters nothing unusual at the horizon.  Perhaps the most conservative resolution is that the infalling observer burns up at the horizon.  Alternatives would seem to require novel dynamics that nevertheless cause notable violations of semiclassical physics at macroscopic distances from the horizon.

\end{abstract}
\end{titlepage}

\baselineskip = 16pt

\tableofcontents

\setcounter{footnote}{0}

\section{Introduction}

The black hole information paradox~\cite{Hawking:1976ra} presents a sharp conflict between quantum theory and general relativity, and so is an important clue to their unification.  Gauge/gravity duality has provided some insight, giving strong evidence that all information is carried away by the Hawking radiation.  It is widely believed that an external observer sees this information emitted by complicated dynamics at or very near the horizon, while an observer falling through the horizon encounters nothing special there.  These three properties --- purity of the Hawking radiation, emission of the information from the horizon, and the absence of drama for the infalling observer --- have in particular been incorporated into the axioms of black hole complementarity (BHC)~\cite{Susskind:1993if,Stephens:1993an}.


Various thought experiments have been examined~\cite{Susskind:1993mu,Preskill}, and argued to show no inconsistency between the observations of the external and infalling observers\footnote{Limits on complementarity with parametrically many fields have been discussed in Ref.~\cite{Yeom:2009zp}.}.  For example, when a bit is thrown into a black hole, then as long as there is a minimum time of order $r_{\rm s} \ln (r_{\rm s}/l_{\rm P})$ before the bit thermalizes and can be reemitted with the Hawking radiation, no observer will see illegal quantum cloning.
This time scale has an interesting resonance with ideas from quantum information theory and from Matrix theory~\cite{Hayden:2007cs,Sekino:2008he,Barbon:2011pn,Lashkari:2011yi}, which suggest that it may actually be achieved.

There would be an inconsistency if one were to consider a large Hilbert space that describes both observers at once.  Such a Hilbert space appears when quantum gravity is treated as an effective field theory, but it cannot be part of the correct theory of quantum gravity if BHC holds.  This is consistent with the idea of holography, wherein quantum gravity is to be constructed in terms of degrees of freedom that are highly nonlocal from the bulk point of view.  For guidance in such uncharted waters, as in the earlier revolutions of relativity and quantum theory, it is important to ask what observations are actually possible.

We will consider first a thought experiment that is a small variation on that of Ref.~\cite{Susskind:1993mu}, differing in that it uses the naturally produced Hawking pairs rather than introducing additional entangled ingoing bits.  This leads us to a rather different conclusion, that the thermalization time does not protect us from an inconsistency of BHC.  Rather, if the experience of the outside observer is as we have assumed, then the infalling observer must encounter high energy quanta at the horizon.  Our first thought experiment requires these only in low partial waves.  However, a second thought experiment, using a detector lowered through the potential barrier to the near-horizon region, allows us to probe higher partial waves and come to the same conclusion about these.  Thus, the infalling observer either burns up at the horizon, or there must be some novel and likely nonlocal dynamics that extends a macroscopic distance from the horizon, as recently proposed in Refs.~\cite{Giddings:2011ks}.  If the latter, we find that the dynamics would have to be of a rather complicated form.

This analysis was inspired in part by the bit models of Refs.~\cite{Mathur:2009hf, Mathur2, Czech:2011wy, Giddings:2011ks, Avery:2011nb, Mathur:2012zp}, and in particular by the theorem that purity of the Hawking radiation implies that the horizon cannot be `information-free'~\cite{Mathur:2009hf} --- that is, unitarity of Hawking evaporation requires $O(1)$ corrections to low energy evolution at the horizon.  We have tried to understand the consequences of this argument for complementarity, and to flesh out the bit model into a more complete picture of the dynamics.   This leads us to a stronger result, namely that there are either order one corrections to the evolution of {\it high energy} modes near the horizon, or the corrections must extend to distances of order the Schwarzschild radius from the black hole.\footnote
{After completion of this work we learned that Ref.~\cite{Braunstein:2009my}  had also argued that black hole evaporation would result in ``a loss of trans-event horizon entanglement'' and thus ``fields far from the vacuum state in the vicinity of the event horizon,'' termed there an {\it energetic curtain.} The argument differs from ours in detail and is based on a model of black hole evaporation that differs from the usual Hawking process and does not satisfy postulate 2. This difference also allows Ref.~\cite{Braunstein:2009my} to conclude that the energetic curtain can be delayed beyond the Page time, possibly even until the black hole evaporates to the Planck scale.}

We note that the three postulates that are in conflict --- purity of the Hawking radiation, absence of infalling drama, and semiclassical behavior outside the horizon --- are widely held even by those who do not explicitly label them as `black hole complementarity.' 

\sect{Complementarity is not enough}
\label{comp}

In considering a slight variant on the thought experiment of Susskind and Thorlacius~\cite{Susskind:1993mu}, we are unable to find an outcome that is consistent with the postulates of black hole complementarity as stated in  Ref.~\cite{Susskind:1993if}:

\begin{quote}

{Postulate 1:} {The process of formation and evaporation of a black hole, as viewed by a distant observer, can be described entirely within the context of standard quantum theory. In particular, there exists a unitary $S$-matrix which describes the evolution from infalling matter to outgoing Hawking-like radiation.}

{Postulate 2:} {Outside the stretched horizon of a massive black hole, physics can be described to good approximation by a set of semi-classical field equations.}

{Postulate 3:} To a distant observer, a black hole appears to be a quantum system with discrete energy levels. The dimension of the subspace of states describing a black hole of mass M is the exponential of the Bekenstein entropy $S(M)$.

\end{quote}
{We take as implicit in postulate 2 that the semi-classical field equations are those of a low energy effective field theory with local Lorentz invariance.}

These postulates do not refer to the experience of an infalling observer, but Ref.~\cite{Susskind:1993if} states a `certainty,' which for uniformity we label as a further postulate:

\begin{quote}

{Postulate 4:} A freely falling observer experiences nothing out of the ordinary when crossing the horizon.
\end{quote}
To be more specific, we will assume that postulate 4 means {both that any low-energy dynamics this observer can probe near his worldline is well-described by familiar Lorentz-invariant effective field theory and also} that the probability for an infalling observer to encounter a quantum with energy $E \gg 1/r_{\rm s}$ (measured in the infalling frame) is suppressed by
an exponentially decreasing adiabatic factor as predicted by quantum field theory in curved spacetime.     We will argue that postulates 1, 2, and 4 are not consistent with one another for a sufficiently old black hole.

Consider a black hole that forms from collapse of some pure state and subsequently decays.  Dividing the Hawking radiation into an early part and a late part, postulate 1 implies that the state of the Hawking radiation is pure,
\begin{equation}
\label{Hstate}
|\Psi\rangle = \sum_i |\psi_i\rangle_E \otimes |i\rangle_L \,.
\end{equation}
Here we have taken
an arbitrary complete basis $|i\rangle_L$ for the late radiation.  Following the ideas of Refs.~\cite{Page:1993df,Hayden:2007cs}, we {use postulates 1, 2, and 3 to} make the division after the Page time when the black hole has emitted half of its initial Bekenstein-Hawking entropy; we will refer to this as an `old' black hole.  The number of states in the early subspace will then be much larger than that in the late
subspace and, as a result,  for typical states $|\Psi\rangle$ the reduced density matrix describing the late-time radiation is close to { the identity}. We can therefore construct operators acting
on the early radiation, whose action on $|\Psi \rangle$ is equal to that of a projection operator onto
any given subspace of the late radiation; this is shown explicitly in
appendix \ref{approx}.

{To simplify the discussion, we treat gray-body factors by taking the transmission coefficients $T$ to have unit magnitude  for a few low partial waves and to vanish for higher partial waves.  A more complete discussion of gray-body factors is included in appendix \ref{gray} and shown to lead to the same basic conclusion that we reach below. Since the total radiated energy is finite, this allows us to think of the Hawking radiation as defining a finite-dimensional Hilbert space. The argument in appendix \ref{approx} assumes that the state of the Hawking radiation is effectively random within this space, as is widely assumed.  We will argue later that this is not necessary.  We also assume, as in Ref.~\cite{Hayden:2007cs}, that the observer knows the initial state of the black hole and also the black hole $S$-matrix.}

Now, consider an outgoing Hawking mode in the later part of the radiation.  We take this mode to be a localized packet with width of order $r_{\rm s}$ corresponding to a superposition of frequencies $O(r_{\rm s}^{-1})$.  {Note that postulate 2 allows us to assign a unique observer-independent lowering operator $b$ to this mode.}   We can project onto eigenspaces of the number operator $b^\dagger b$.  In other words, an observer making measurements on the early radiation can know the number of photons that will be present in a given mode of the late radiation.

Following postulate 2, we can now relate this Hawking mode to one at earlier times, as long as we stay  outside the stretched horizon.  The earlier mode is blue-shifted, and so may have frequency $\omega_*$ much larger than $O(r_{\rm s}^{-1})$ though still sub-Planckian.

Next consider an infalling observer and the associated set of infalling modes with lowering operators $a$.  Recall that Hawking radiation arises precisely because
\begin{equation}
\label{bog}
b =  \int_0^\infty d\omega \left(B(\omega) a_\omega +  C(\omega)  a^\dagger_\omega
 \right)\,,
\end{equation}
so that the full state  cannot be
  both an $a$-vacuum ($a  |\Psi\rangle =0$) and a $b^\dagger b$ eigenstate.  Here we have again used our simplified gray-body factors.

The application of postulates 1 and 2 has {thus} led to the conclusion that the infalling
observer will encounter high-energy modes.
Note that the infalling observer need not have actually made the measurement on the early radiation:  to guarantee the presence of the high energy quanta it is enough that it is possible, just as shining light on a two-slit experiment destroys the fringes even if we do not observe the scattered light.
{Here we make the implicit assumption that the measurements of the infalling observer can be described in terms of an effective quantum field theory.  Instead we could simply suppose that if he chooses to measure $b^\dagger b$ he finds the expected eigenvalue, while if he measures the noncommuting operator $a^\dagger a$ instead he finds the expected vanishing value.  But this would be an extreme modification of the quantum mechanics of the observer, and does not seem plausible.}

Fig.~1 gives a pictorial summary of our argument, using ingoing Eddington-Finkelstein coordinates.
\begin{figure}[t]
\begin{center}
\includegraphics[scale=.8]{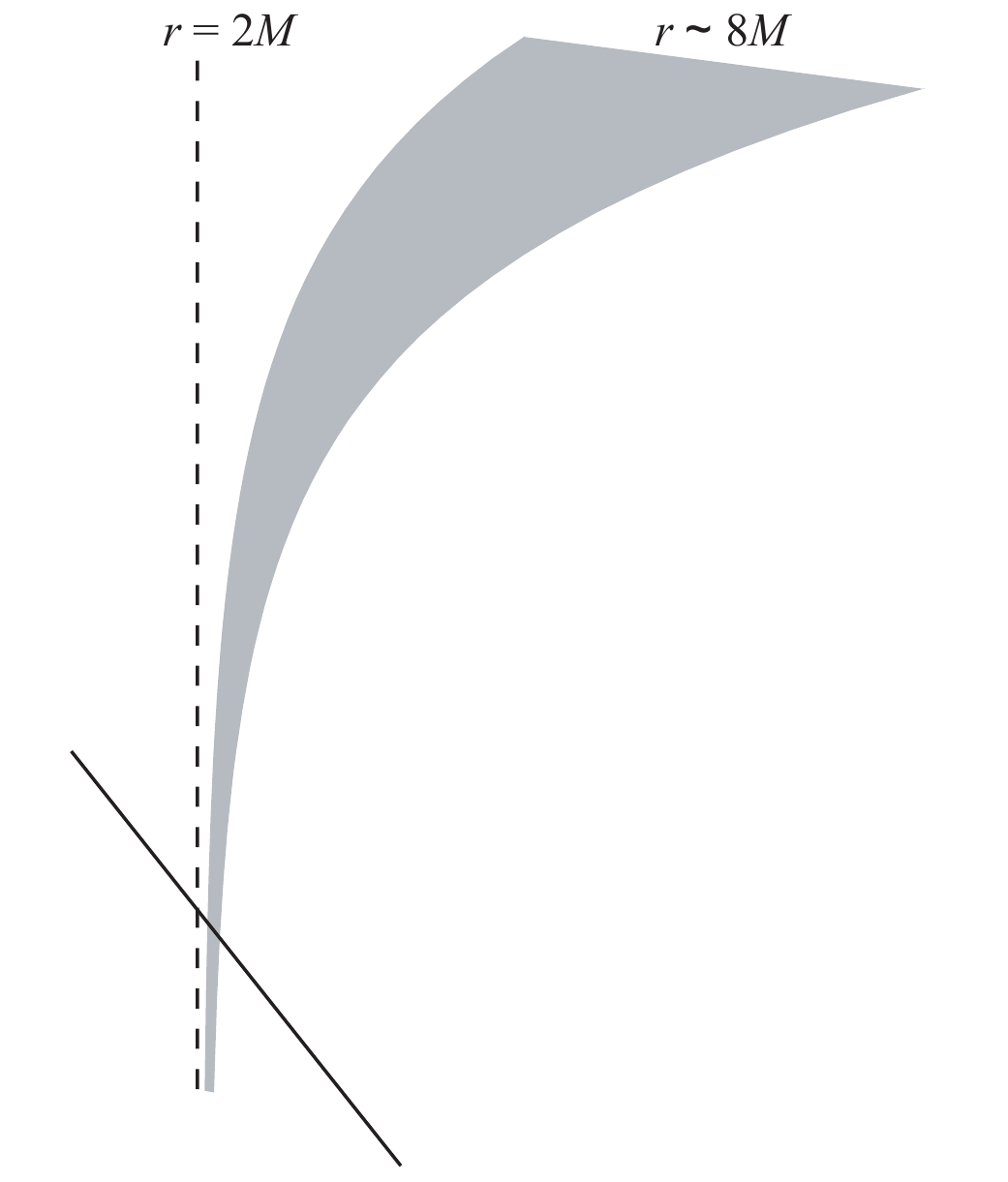}
\end{center}
\vspace{-10pt}
\caption{Eddington-Finkelstein coordinates, showing the infalling observer encountering the outgoing Hawking mode (shaded) at a time when its size is $\omega_*^{-1} \ll r_{\rm s}$.   If the observer's measurements are given by an eigenstate of $a^\dagger a$, postulate 1 is violated; if they are given by an eigenstate of $b^\dagger b$, postulate 4 is violated; if the result depends on when the observer falls in, postulate 2 is violated.}
\end{figure}
The support of the mode $b$ is shaded.  At large distance it is a well-defined Hawking photon, in a predicted eigenstate of $b^\dagger b$ by postulate 1.  The observer encounters it when its wavelength is much shorter: the field must be in the ground state $a_\omega^\dagger a_\omega = 0$, by postulate 4, and so cannot be in an eigenstate of $b^\dagger b$. But by postulate 2, the evolution of the mode outside the horizon is essentially free, so this is a contradiction.

To restate our paradox in brief, the purity of the Hawking radiation implies that the late radiation is fully entangled with the early radiation, and the absence of drama for the infalling observer implies that it is fully entangled with the modes behind the horizon.  This is tantamount to cloning.  For example~\cite{Mathur:2009hf}, it violates strong subadditivity of the entropy,
\begin{equation}
S_{AB} + S_{BC} \geq S_B + S_{ABC}\,.
\end{equation}
Let $A$ be the early Hawking modes, $B$ be our outgoing Hawking mode, and $C$ be its interior partner mode.  For an old black hole, the entropy is decreasing and so $S_{AB} < S_A$.  The absence of infalling drama means that $S_{BC}=0$ and so $S_{ABC} = S_A$.
{Subadditivity then} gives $S_A \geq S_B + S_A$, which fails substantially since the density matrix for system $B$ by itself is
thermal.  {This argument is exactly as in Ref.~\cite{Mathur:2009hf}, where the additional observations we are making are that i) the reasoning holds even under the postulates of black hole complementarity and ii) the modes $b$ and $c$ have high energy as seen by an infalling observer.}

Actually, assuming the Page argument~\cite{Page:1993df}, the inequality is violated even more strongly: for an old black hole the entropy decrease is maximal, $S_{AB} = S_A - S_B$, so that we get from subadditivity that $S_A \geq 2S_B + S_A$.  Appendix \ref{approx} makes an equivalent assumption, the randomness of the Hawking state, in order to show that  measurements of the early radiation
predict the state of the late mode with high fidelity.  We see from the subadditivity argument that this strong assumption is not
needed; it is sufficient that the entropy of the black hole be decreasing.  From another point of view,
one need not be able to predict the state with perfect fidelity; rather, any information about the state of the $b$ mode precludes the state being annihilated by $a$.

Note that the measurement of $N_b$ takes place entirely outside the horizon, while the measurement of $N_a$ (real excitations above the infalling vacuum) must involve a region that extends over both sides of the horizon.  These are noncommuting measurements, but by measuring $N_b$ the observer can infer something about what would have happened if $N_a$ had been measured {instead}.  For an analogy, consider a set of identically prepared spins.  If each is measured along the $x$-axis and found to be $+\frac12$, we can infer that a measurement along the $z$-axis would have had equal probability to return $+\frac12$ and $-\frac12$.  The multiple spins are needed to reduce statistical variance; similarly in our case the observer would need to measure several modes $N_b$ to have confidence that he was actually entangled with the early radiation.

One might ask if there could be a possible loophole in the argument: A physical observer will have a nonzero mass, and so the mass and entropy of the black hole will increase after he falls in.  However, we may choose to
consider a particular Hawking wavepacket which is already separated from the streched horizon by a finite amount when it is encountered by the infalling observer.  Thus by postulate 2 the further evolution of this mode is semiclassical and not affected by the subsequent merging of the observer with the black hole.  In making this argument we are also assuming that the dynamics of the stretched horizon is causal.

Ref.~\cite{Nomura:2012sw}, in response our argument, has claimed that it is not possible to measure the state of the early radiation in the basis that is required for our argument due to gravitational effects.  Essentially we are taking the early radiation as input to a quantum computation in the flat region distant from the black hole, which then returns the desired bit in an easily measured form.   We do not see an argument that would forbid such computations.
Note that in order to distinguish a pure state of Hawking radiation from a mixed state, it is also necessary to measure the state of the radiation in many bases; thus the claim of Ref.~\cite{Nomura:2012sw} would mean that there is no information problem in the first place.

Ref.~\cite{Suss2} raises a related issue, that the quantum computation might take longer than the black hole lifetime to carry out.  The most precise formulation of the information problem is in AdS spacetime, where the geometry itself serves to confine the Hawking quanta~\cite{Hawking:1982dh,Maldacena:2001kr}.  We can use a similar strategy here: consider an ${\cal N}=4$ Yang-Mills theory on $S^3$, dual to AdS gravity, in addition to a reference system consisting of a large collection of spins.  Prepare the total system in a pure state, with the Yang-Mills theory fully entangled with the spins, such that its density matrix is thermal with a temperature above the Hawking-Page transition~\cite{Hawking:1982dh}.  In effect we are using the spins in place of the early radiation.  {But since the spins do not live in the AdS space, there are clearly no constraints of time or gravitation that could} prevent an observer from measuring $\hat P$ and then diving into the black hole and finding, via our argument, high energy quanta.\footnote{
We have deliberately chosen the reference system not to have a geometric dual.  In the case where it is a second copy of the gauge theory~\cite{Maldacena:2001kr}, there are additional subtleties~\cite{MarWall}.}

Thus far the {asymptotically flat} discussion applies to a black hole that is older than the Page time; we needed this in order to frame a sharp paradox using the entanglement with the Hawking radiation.  However, we are discussing what should be intrinsic properties of the black hole, not dependent on its entanglement with some external system.  After the black hole scrambling time~\cite{Hayden:2007cs,Sekino:2008he}, almost every small subsystem of the black hole
is in an almost maximally mixed state. So if the degrees of freedom sampled by the infalling observer can be considered typical, then they are `old' in an
intrinsic sense.  Our conclusions should then hold.  If the black hole is a fast scrambler the scrambling time is $r_{\rm s} \ln (r_{\rm s}/l_{\rm P})$, after which we have to expect either drama for the infalling observer or novel physics outside the black hole.   Ref.~\cite{Suss2} has suggested that the existence of the high energy quanta might be a special observable which is governed by the Page time rather than the fast scrambling time.  We view this as an open question {pending development of a dynamical theory of these quanta and how they form.}


\sect{Further discussion}

\label{disc}

\subsection{Extension to higher partial waves}

It is well known that Hawking radiation from an asymptotically flat Schwarzschild black hole is dominated by low angular momentum modes; see e.g. \cite{Page:1976df}.  This is a consequence of the fact that a black hole of  Hawking temperaure $T_H$ and Schwarzschild radius $r_s$ has $T_H  r_s \sim 1$, so that high angular momentum modes of energy $T_H$ are trapped behind a large barrier in the effective radial potential.  One might therefore be tempted to believe that the issue discussed in section \ref{comp} concerns only a small number of partial waves.  Since a local observer is unlikely to encounter such quanta, one might then conclude that a (much-weakened) version of postulate 4 might still hold in which the suppression is replaced by a fixed (1/area) power law.\footnote{In addition, one would need to propose a mechanism through which these quanta would arise from the infalling perspective.  This would appear to require that the infalling observer experience violations of local quantum field theory at this (power-law-suppressed) level.}

This would already be a striking result: these quanta must appear quite close to the horizon (see Fig.~1) and so violate the standard wisdom that the horizon is not a distinguished location.  And they are not rare {in the sense that} their number is of the same order as the number of actual Hawking quanta. However, we will argue for an even stronger result, by considering a thought experiment in which the centrifugal barrier is penetrated.

As
noted long ago by Unruh and Wald \cite{Unruh:1982ic},
it is possible to `mine' energy from the modes trapped behind the effective potential.  The basic procedure is to lower some object below the potential barrier, let the object absorb the trapped modes, and then raise the object back above the barrier.  Unruh and Wald thought of the object as a box that could be opened to collect ambient radiation and then closed to keep the radiation from escaping.  One may also visualize the object as a particle detector, though the two are equivalent at the level discussed here.

We analyze a particular version of the mining process in appendix \ref{mining} in order to address gravitational back-reaction and other concerns not considered in \cite{Unruh:1982ic}; see also \cite{AB} for similar conclusions. While these additional issues limit the rate at which our process can mine energy to below that predicted by \cite{Unruh:1982ic} (see footnote \ref{lim}), they do not change the basic result that energy can be extracted from the high angular momentum modes.  In fact, we are unable to identify any fundamental constraint that would forbid the extraction of energy from any mode separated from the horizon by more than a Planck distance $\ell_p$.

In the context of such a mining operation, the arguments of section \ref{comp} can be applied to the higher partial waves as well.  One need only consider the internal state of the mining equipment to be part of the late-time Hawking radiation.  In particular, postulate 2 can be used to evolve the mode $b$ to be mined backward in time and to conclude for an old black hole that, even before the mining process takes place, the mode must be fully entangled with the early-time radiation.  Postulate 4 is then violated for these modes as well, suggesting that the infalling observer encounters a Planck density of Planck scale radiation and burns up.  One might say that the black hole is protected by a Planck-scale firewall.

The arguments that we have given for the firewall are largely built on those {that have been used} to support the fuzzball picture~\cite{Mathur:2009hf}, and one might wonder whether the fuzzball provides the actual dynamical picture of the firewall.  It is not clear to what extent there is a well-defined fuzzball construction for macroscopic nonextremal black holes, {but our conclusions seem to contradict the scenario advocated in}~\cite{Mathur:2012zp}, which incorporates a form of complementarity such that an infalling observer sees nothing unusual on the horizon (though he may be constructed in some dual form on the branes).  Since the branes are thought of as extending only a microscopic
 distance above the horizon, essentially a realization of the stretched horizon, postulate 2 holds and our argument would apply.   Ref.~\cite{Mathur:2012jk} has reiterated this complementarity conjecture, but does not address our arguments directly.

A sharp end to spacetime, similar to the firewall, arose in a related context in Ref.~\cite{Czech:2012be} {(though see \cite{MarWall} for comments)}.  Additional earlier suggestions that the
geometry end at the horizon, or that the interior geometry is very different from Schwarzschild, include Ref.~\cite{Chapline:2000en}, whose connection with general relativity is not clear, Ref.~\cite{Mazur:2001fv}, using sources that violate various energy conditions, and Ref.~\cite{Davidson:2011eu}, using a higher derivative action.  Other works have reached similar conclusions from quite different starting points. These include Ref.~\cite{Itzhaki:1996jt}, which attempts to argue that the black hole S-matrix hypothesis requires strong interactions between ingoing and outgoing particles such that the former never cross the horizon, and ref. \cite{Lowe:2006xm}, which starts from the assumption that acts far outside the black hole can causally affect spacelike separated observers in the interior.

Note that this firewall need not be visible to any observer that remains outside the horizon.  All that we have argued is that the infalling observer does not experience a pure state.  There remains considerable freedom in the possible reduced density matrices that could describe a few localized degrees of freedom outside the black hole, so that this matrix might still agree perfectly with that predicted by Hawking \cite{Hawking:1974sw}.  In this case any local signal that an external observer might hope to ascribe to the firewall at distance $1/\omega_*$ cannot be disentangled from the Unruh radiation that results from probing this scale without falling into the black hole.

\subsection{Relaxing postulate 2?}

Postulate 2 plays a crucial role in any version of our argument, allowing us to use low energy effective gravity to {associate a unique observer-independent operator $b$ with the designated mode and to evolve it in time.}  The purity of Hawking radiation implies a breakdown of semiclassical physics, but the usual complementarity assumption as stated in this postulate is that the complicated dynamics that leads to re-emission of information takes place, from the point of view of the exterior observer, only on the stretched horizon a Planckian distance above the event horizon.  A possible alternative to the firewall is thus that this postulate should be relaxed, giving some
novel (and perhaps non-local) evolution that extends a finite distance from black hole as has recently been proposed in Ref.~\cite{Giddings:2011ks}.
We agree with \cite{Giddings:2011ks} that one would like to keep such novel physics to a minimum.

However, if we are to relax postulate 2 then the {modified} dynamics must not only be more nonlocal than expected, but also much larger in magnitude.  It is generally believed that the return of information requires modification of the Hawking calculation only for observables involving $O(S)$ quanta, or in effects of order $e^{-S}$, or~\cite{Maldacena:2001kr} for small numbers of quanta over extremely long time-scales.  However, preservation of postulates 1 and 4 requires that an $N_a$ eigenstate evolve to an $N_b$ eigenstate, which is an $O(1)$ effect visible in the two-point function over time scales not much larger than the light-crossing time.

Note that
{our thought experiment} is very similar to that in Ref.~\cite{Susskind:1993mu}, except that instead of using bits thrown into the black hole, it uses the naturally produced Hawking bits.  In the former case, an observer who has seen the exterior bit cannot see its interior clone, basically because it is too deep in the interior after a scrambling time of at least $r_{\rm s} \ln (r_{\rm s}/l_{\rm P})$.  In the case we consider, the scrambling time does not seem to enter in the same way: the infalling observer encounters the high energy quantum right behind the horizon, at a distance $\omega_*^{-1}$.\footnote{It is worth noting that nowhere in our argument do we consider `nice slices,' which extend deep into the black hole interior and which often enter into discussions of the breakdown of effective field theory in black holes.  All observations are limited to the exterior and a small distance $\omega_*^{-1}$ behind the horizon.}

We should therefore ask how the scrambling time might affect the argument.  Decay is not an equilibrium process, and after emission of a Hawking quantum there will be a delay before the black hole returns to its typical state, just as there is when it absorbs a quantum.\footnote{To be precise, it never reaches a fully typical state, as additional emissions occur in the meantime.}  It is interesting that the conjectured fast-scrambling time $r_{\rm s} \ln (r_{\rm s}/l_{\rm P})$ is the same magnitude as the time during which the Hawking mode moves out from the stretched horizon to a macroscopic distance $O(2 r_{\rm s})$, and during which it redshifts from a near-Planckian energy to $O(r_{\rm s}^{-1})$.  We therefore investigate what form of time evolution would be needed to restore postulate 4.

Consider an old black hole containing $N$ bits in a basis state $|j\rangle$; the full state of the system is given by a sum over $j$, entangled with the outgoing Hawking radiation.  Immediately after emission of a Hawking mode (which we idealize as a single bit) from the stretched horizon, postulate 4 requires that the mode be entangled with the modes behind the horizon.  We must therefore use a state of $N+1$ bits to describe the resulting hole, so that the evolution is
\begin{equation}
\label{vacuum}
|j\rangle \to \sum_k |j , k ; k\rangle \,.
\end{equation}
We have taken a convenient basis in which $k$ is the state of the Hawking bit and we have singled out the interior bit with which it is entangled.  After the thermalization time, the hole has only $N-1$ bits, and
\begin{equation}
\sum_k |j , k ; k\rangle \to \sum_{l,m} |l; m\rangle \langle l;m|j\rangle \,. \label{therm}
\end{equation}
The $N$ bits of $j$ are mapped into the $N-1$ bits of $l$ plus the outgoing bit $m$.
The effect is that one bit of entanglement with the earlier radiation is transferred to to the outgoing bit $k$.

Eq.~(\ref{therm}) describes unitary evolution from an $N$ bit space indexed by $j$ to $(N-1)+1$ bit spaces indexed by $l$ and $k$.  The state on the left is embedded in a space of $N+2$ bits, but the evolution has been specified only when two are in a definite state.  For any other state of these two bits there is a high energy quantum near the horizon, {which} should be atypical in the black hole Hilbert space.  Our description differs from the bit models of \cite{Mathur:2009hf, Czech:2011wy, Giddings:2011ks, Avery:2011nb, Mathur:2012zp} through the explicit description of these bits before thermalization occurs,  i.e.\ the intermediate state in Eqs.~(\ref{vacuum}, \ref{therm}).  This will play a key role below.  Note that the evolution~(\ref{therm}) cannot be thought of as simple thermalization of the black hole, because it evolves from a Hilbert space of $N+1$ bits to one of $N-1$ bits.\footnote{We could try to describe this as acting only on the black hole degrees of freedom by projecting with a Hawking mode  $\langle k' |$ to get
\begin{equation}
\label{new}
|j , k' \rangle \to \sum_l  |l \rangle \langle l;k'|j\rangle \,.
\end{equation}
This is nonunitary evolution from a space of $N+1$ bits to one of $N-1$ bits: note that $k'$ is in the ket on the left and the bra on the right.  This is similar to the nonunitary evolution appearing in the black hole final state conjecture~\cite{Horowitz:2003he}.}
Rather, it acts unitarily on the whole \{black hole + outside Hawking mode\} system.

In other words, we have again arrived at the above-mentioned possibility that novel and perhaps non-local dynamics extends a finite distance $\ell_{new}$ from the black hole.  The size of $\ell_{new}$ will set the scale of radiation encountered by the infalling observer.  If this novel physics is associated with thermalization, then $\ell_{new} \sim r_s$ as proposed in \cite{Giddings:2011ks} so that an infalling observer sees only radiation with $\omega_* \sim r_s^{-1}$ in rough agreement with the prediction of local field theory.

Perhaps this is the way things work though, if so, there seem to be significant further implications.  Ref \cite{Giddings:2011ks} envisioned this new effect as acting only on a few partial waves of otherwise essentially free fields.  The analogous statement here would be that it acts on the internal state of any mining equipment used to extract energy from the black hole, including for example
notes that the equipment might print on paper and then lock in a vault in order to record the results of the experiment.

Even this appears appears to be insufficient.
{Let us} suppose that the equipment can manipulate the quantum data in the storage bit, say on receipt of a signal from far away, so as to perform an arbitrary unitary $U$ transformation on the
 storage bit.  Then the analogue of equation \eqref{vacuum} becomes
\begin{equation}
\label{transform}
|j\rangle \to \sum_k |j , k ; U k\rangle \,.
\end{equation}
We might take $U$ to permute the storage bit basis states $k$, or we might take it to act as the phase $(-1)^k$.
For each $|j\rangle$, allowing $U$ to range over all unitary operations generates a basis for a Hilbert space of dimension $4$. In this sense, the right hand side of \eqref{transform} spans a full $N+2$ bit Hilbert space. There can thus be no $U$-independent analogue of equation \eqref{therm} involving only a remaining $N-1$ bit black hole and 1 additional storage bit.  {Note that explicit dependence of the Hamiltonian on $U$ would violate the usual rules of quantum mechanics.}

Unless there is some physical constraint that restricts their initial state,
including any other finite number of bits is not helpful.   Without such a restriction, these bits can neither provide a useful record of transformation $U$, nor can they be used as an empty box into which to deposit the information about $U$ in \eqref{transform}.   They simply add equally to the dimensions of the Hilbert spaces on the left- and right-hand sides of the supposed new version of \eqref{therm} with no effect on the 2 bit mismatch noted above.

Since one clear restriction is the existence of the storage bits themselves, an effect that destroys these bits as they are transported back to large $r$ might suffice. A final alternative might be to couple
to the infinite number of states associated with occupation numbers in outgoing radiative modes, though one would expect such a coupling to modify even the mean rate at which energy and/or information escape from the black hole.  Seeing no more gentle alternatives, we therefore disagree with \cite{Giddings:2011ks} that this new physics can be `innocuous' in all of the senses described there.

The alternative would appear to be that some yet unknown new physics (or some effect that we have neglected) simply prevents energy from being mined closer to the horizon than $\ell_{new}$. This might be a new fixed scale or some geometric mean of $\ell_p$ and $r_s$. There would then be no obvious reason to believe that infalling observers experience radiation above the scale $1/\ell_{new}$, though one would certainly expect them to see {\it some} violation of local quantum field theory.  This scenario is realized in certain models of local quantum field theory that violate local lorentz invariance \cite{breakL}.  In these models, there are simply no outgoing modes within distance $\ell_{new}$ of the horizon.  Some additional physics would of course be needed to transfer information to the Hawking-like modes at the scale $\ell_s$, but since we have removed the possibility of mining the radiation at a lower scale,  this effect can now be limited to the natural Hawking-like modes themselves.

Recently, Refs.~\cite{Banks,Srednicki,Bousso:2012as,Harlow:2012me} have suggested that an extended {``strong"} complementarity principle might survive.  In particular, an asymptotic observer would see an eigenstate of $N_b$, and infalling observer an eigenstate of $N_a$.  This abandons the strong form of postulate 2, that we can use effective field theory freely outside the horizon, but has been argued to preserve a weaker postulate that no single observer can see a violation of effective field theory.  We disagree.  One can consider a continuous family of possible observers, falling in at different times.  One who meets a mode near the stretched horizon sees an $N_a$ eigenstate, and one who meets it far from the black hole sees an $N_b$ eigenstate.  Observers in between would see a continuum of interpolating states.  But two close-spaced observers can communicate (or a single observer can carry an apparatus that measures time-dependence near his world-line), so this time-dependence is detectable and violates effective field theory: the weakened postulate {appears to be} no safer than the strong one.

More theoretically, note that Eq.~(\ref{bog}) is basic to the derivation of Hawking radiation, with the LHS evaluated {by an asymptotic observer and the RHS constrained by the fact that an infalling observer sees near-vacuum (by the adiabatic principle).} If one rejects this as meaningless because no observer can see both sides, then one has the burden of providing a new theory to derive the Hawking flux, in which this equation is replaced by something presumably more complicated.  {Still further concerns about this idea were raised in \cite{Chowdhury:2012tr}.}

\sect{Conclusions}

Historically, the black hole information paradox presented three main alternatives, each problematic: information loss, purity of the Hawking radiation, and remnants.  The discovery of gauge/gravity duality pointed to purity, and to a fundamentally nonlocal formulation of quantum gravity.  Our work again seems to present some sharp and perhaps unpalatable alternatives: a firewall at the horizon, or novel and probably nonlocal dynamics extending a macroscopic distance outside the horizon.  (We note that the firewall also has elements of nonlocality, in that its location, the horizon, is not determined by any local feature but by a global property.)  The second alternative has the potential to connect with one of the notable features of BHC, the fast-scrambling time scale, but our attempt to determine possible forms of the dynamics leads us to conclude that it would {nevertheless cause notable violations of semiclassical physics at macroscopic distances from the horizon.}

The tensions noted in this work may lead the reader to wonder whether even the most basic coarse-grained properties of Hawking emission as derived in \cite{Hawking:1974sw} are to be trusted.  But the thermodynamic picture of black holes now rests on many pillars that remain intact.  Even at the microscopic level, at least in string theory, independent evidence for thermal emission from black holes  comes from studies of low energy excitations of D-branes and from AdS/CFT.  This leads one to suspect that some appropriately weakened version of postulates 2 and 4 might be retained in order to help explain the success of the Hawking calculation, though finding a consistent scenario remains a challenge.

Let us conclude by briefly commenting on more general causal horizons, which of course share many features in common with black holes.  For example, the reader will note that we pass through Rindler horizons all the time and do not burn up or experience obvious new physics.   We believe that this is due to an essential difference between Rindler and black hole horizons.
Since Rindler horizons have infinite entropy, their quantum memory never fills.  `Young' Rindler horizons never evolve to become `old.' From another point of view, the fact that Rindler horizons do not evaporate makes it impossible to apply the arguments of section \ref{comp}.

One might also ask about cosmological horizons, such as those in de Sitter space.  These are more like black holes in that they have finite entropy, though they still do not evaporate.  The experimental evidence is also not clear cut.  Our present universe is just now emerging into an era dominated by dark energy.  As a result, any cosmological horizons through which we cross soon should be expected to be young.  Even if they behave like fast-scrambling ($r_s \ln r_s$) black holes it will be a time $\sim 60$ times the age of our universe before they become old.  On the other hand, the fact that early universe inflation must last more than 60 $e$-foldings suggests that the associated cosmological horizons may have become old.  We leave for future work the question of whether this would significantly affect its predictions for cosmology and whether this argues that, despite their finite entropy, cosmological horizons differ fundamentally from black holes.

\section*{Acknowledgments}
We thank Raphael Bousso, Adam Brown, Steve Giddings, David Gross, Daniel Harlow, Patrick Hayden, Samir Mathur, Yasunori Nomura, John Preskill, Mark Srednicki, Douglas Stanford,  Lenny Susskind, Bill Unruh, Aron Wall, and all of the participants of the KITP Bits, Branes, and Black Holes program for useful discussions.
AA, JS, and JP were supported in part by NSF grants PHY05-51164 and PHY07-57035, and by FQXi grant RFP3-1017.  DM was supported in
part by the National Science Foundation under Grant Nos PHY11-25915
and PHY08-55415, by FQXi grant RFP3-1008, and by funds from the University of California.  He also thanks the Kavli Institute for Theoretical Physics for their hospitality during much of this work.

\appendix

\sect{Approximate projection operators}
\label{approx}

Consider the projection operator onto state $|i\rangle_L$ in some orthonormal basis for the late radiation, $P^i = |i\rangle_L \langle i |_L$.
We consider the case that the Hawking state $|\Psi\rangle$ is chosen with uniform measure, as in the microcanonical ensemble; in Appendix B we will discuss a slight generalization.  Then
 the operator
\begin{equation}
\hat P^i = L |\psi_i\rangle_E \langle \psi_i |_E \,,
\end{equation}
which acts on the state of the early radiation, allows us to anticipate the measurement of $P^i$ if $E \gg L$.  Here $E$ and $L$ are the dimensions of the early and late Hilbert spaces (so $1 \leq i,j,\ldots \leq L$, while $1 \leq a,b,\ldots \leq E$ for an $E$-basis to be introduced later).  That is,
\begin{equation}
\hat P^i  |\Psi\rangle \approx P^i |\Psi\rangle =  |\psi_i\rangle_E \otimes |i\rangle_L  \,.
\end{equation}
If the $ |\psi_i\rangle_E$ were orthogonal with equal norms, this would be an equality, and we show that it approaches this for typical states $|\Psi\rangle$ when $L \gg E$.

The relative error is
\begin{equation}
{\cal E} = \frac{ \| ( P^i -  \hat P^i) |\Psi\rangle \|^2}{ \| P^i |\Psi\rangle \|^2 } = \left(1 - L \langle  \psi_i |  \psi_i \rangle_E \right)^2
+ L^2 \sum_{ j \neq i} |\langle  \psi_i |  \psi_j \rangle_E|^2
\end{equation}
Expanding in an orthonormal basis $|\psi_i \rangle_E = \sum_{a=1}^E c_{ia} | a\rangle_E$, the average over
$|\Psi\rangle$ with the uniform measure gives
\begin{equation}
\overline{  c_{ia}  c^*_{jb} } = \frac{1}{LE}{\delta_{ij} \delta_{ab}}  \,, \qquad \overline{  c_{ia}  c^*_{jb} c_{kc} c^*_{ld} } =
\frac{1}{L^2 E^2} (\delta_{ij} \delta_{kl} \delta_{ab} \delta_{cd} + \delta_{il} \delta_{jk} \delta_{ad} \delta_{bc}  )
\label{uniform}
\end{equation}
(dropping terms of relative order $1/LE$), and so
\begin{equation}
\overline{  \langle  \psi_i |  \psi_j\rangle_E } = \frac{1}{L}{\delta_{ij} }  \,, \qquad \overline{   \langle  \psi_i |  \psi_j\rangle_E   \langle  \psi_k |  \psi_l\rangle_E  } =
\frac{1}{L^2}  \delta_{ij} \delta_{kl}  + \frac{1}{L^2 E} \delta_{il} \delta_{jk}  \label{uniform2}\,.
\end{equation}
Then for $E \gg L \gg 1$,
\begin{equation}
\label{ebar}
\overline{ \cal E  } =
\frac{L}{E}  \,.
\end{equation}
This decreases exponentially after the halfway point of the black hole's life.  While the explicit calculations above refer to projections onto a one-dimensional space, \eqref{ebar} also holds for more general projections given by sums of the $\hat P^i$ above.

\sect{Effects of Gray-body factors}
\label{gray}

In the linear approximation, each quantum field outside the black hole may be decomposed using spherical harmonics.  Each mode then leads to an effective 1+1 scattering problem in an effective potential which depends on the mode's angular momentum $j$.  The annihilation operators $b,c,d$ corresponding to the outgoing mode outside the barrier ($b$), the incoming mode outside the barrier ($c$), and the outgoing mode inside the barrier ($d$) are then related by reflection and transmission coefficients $R,T$ through $b = T^*d + \frac{RT^*}{T}c$, so that
\begin{equation}
N_b = |T|^2 N_d + RT^* d^\dagger c + R^* T c^\dagger d + |R|^2 N_c.
\end{equation}

On the other hand,  \eqref{bog} now becomes
\begin{equation}
\label{bog2}
d =  \int_0^\infty d\omega \left(B(\omega) a_\omega +  C(\omega)  a^\dagger_\omega
\right)\, .
\end{equation}
As usual in a scattering problem, the incoming modes on opposite sides of the barrier are completely independent.  Thus $c, c^\dagger$ commute with $a_\omega, a_\omega^\dagger$.

Although the gray-body coefficients complicate the relation between the outgoing Hawking modes $b$ and the infalling modes $a$, it remains true that the number operators $N_b,N_{a_\omega}$ fail to commute unless the transmission coefficient $T$ is very small.  In particular, even when acting on a state in the $a,\tilde a$, and $c$ vacuum ($c |\psi \rangle = a_\omega |\psi \rangle = \tilde a_\omega |\psi \rangle$ for all $\omega$) we have

\begin{equation}
\label{flux}
N_b |\psi \rangle = T(T^*d^\dagger + R^* c^\dagger) \int_0^\infty d\omega \, C(\omega)  a^\dagger_\omega |\psi \rangle,
\end{equation}
which for an infalling observer contains of order $|T|^2$ particles for small $T$.

Since $T$ decreases exponentially for large $j$, the state \eqref{flux} is indistinguishable from the infalling vacuum for large $j$.  But for the first few partial waves it leads to a noticeable flux of particles for infalling observers.

 Due to gray body factors, the state of the Hawking radiation also deviates from the microcanonical ensemble assumed in Appendix A. To model
 this effect we replace
\begin{equation}
\delta_{ij}/L \to p_j \delta_{ij}\,,\quad \delta_{ab}/E \to \tilde p_a \delta_{ab}
\end{equation}
 in the expectation values~(\ref{uniform}), with the $p_i$ and $\tilde p_a$ each summing to unity.  Then the expectation values~(\ref{uniform2}) become
\begin{equation}
\overline{  \langle  \psi_i |  \psi_j\rangle_E } = p_i {\delta_{ij} }  \,, \qquad \overline{   \langle  \psi_i |  \psi_j\rangle_E   \langle  \psi_k |  \psi_l\rangle_E  } =
p_i p_k \delta_{ij} \delta_{kl}  + p_i p_k ( {\textstyle \sum_{a}  \tilde p_a^2} ) \delta_{il} \delta_{jk}\,.
\end{equation}
For the approximate projection operator we take
\begin{equation}
\hat P^i = |\psi_i\rangle_E \langle \psi_i |_E/p_i \,.
\end{equation}
One then finds
\begin{equation}
\overline{ \cal E  } =
\frac{ \textstyle\sum_{a}  \tilde p_a^2}{p_i }  \,.
\end{equation}
The numerator is of order $1/E'$ where $E'$ is the number of states of the early radiation that are populated with significant probability.  The denominator is similarly of order $1/L'$ for states of the final radiation that are populated with significant probability, so the conclusion is the same as before for these states.  Note that we have assumed that the late-time density matrix is diagonal in the late basis $i$ in which we project.  The semi-classical analysis \cite{Hawking:1974sw} suggests that this is the case, to good approximation, for the occupation number basis.


\sect{Black hole mining revisited}
\label{mining}

We now study a specific process for mining energy from the high angular momentum modes of a (say, Schwarzschild) black hole's thermal atmosphere in order to examine constraints beyond those addressed in \cite{Unruh:1982ic}.  These modes lie close to the horizon.  We therefore wish to lower a detector to within a proper distance $L \ll r_s$ of the horizon, so that we probe modes of angular momentum $r_s/L \gg 1$. We take the detector to be of size $\sim \epsilon L$ and (unexcited) mass $m_{\rm det} = \epsilon^{-1} L^{-1}$. Here $\epsilon$ is a small constant (e.g., $1/100$ or $10^{-6}$) independent of $r_s$ and the Planck scale $\ell_p$.  The detector is attached to one end of a tension $\mu$ cosmic string.\footnote{The extraction of energy from black holes via cosmic strings was also studied by Lawrence and Martinec \cite{LM} and by Frolov and Fursaev \cite{FF}.  They considered strings that pierce the horizon, while we intentionally keep our apparatus outside.  We avoid direct coupling to the black hole so as not to confuse our investigation of the high angular momentum modes in the thermal atmosphere of the black hole.}   This attachment presumably makes use of an appropriate monopole that allows the string to end, which we think of as part of our detector.  The other end of the cosmic string is attached to a static Dyson sphere of radius $r_0 \sim \epsilon^{-1} r_s$ which completely encloses the black hole.\footnote{We choose a Dyson sphere for simplicity.  One could also use orbiting space stations.  For small enough orbital velocities, the motion of the space station should not affect the detector during the time that it is active.}  We work in $d \ge 4$ spacetime dimensions.

The detector is to be lowered from $r_0$ to within a proper distance $L \gg \ell_p$ of the horizon, where the locally measured temperature is $T_{\rm loc} \sim 1/L$.   Since  $m_{\rm det} \gg T_{\rm loc}$, the detector can remain stable in this thermal bath.  In particular, there is little danger of it colliding with an anti-detector in black hole's thermal atmosphere.

At higher altitudes the detector does not interact significantly with the black hole's thermal atmosphere due to its small physical size and the resulting small cross-section for absorption.  But it will begin to do so at the target height $L$.  The absorption of a Hawking photon increases the mass of the detector by  the relatively small amount $T_{\rm loc} \sim L^{-1} = \epsilon m_{\rm det}$.  The detector is to be left in place long enough to absorb a Hawking photon (which requires an asymptotically measured time of order $r_s$) and then lifted back to $r_0$.  As discussed in \cite{Unruh:1982ic}, the net amount of energy extracted from the black hole is of order $T_H$.  We must choose $\mu =
m_{\rm det}/L \sim \epsilon^{-1} L^{-2}$ so that it can support the weight of the detector at the height $L$.    This condition also ensures that the local temperature at $L$ satisfies $T_{\rm loc}^2 \ll \mu$ so that closed loops of string are not a significant part of the thermal atmosphere at this depth.  We note that the natural width $\mu^{-1/2}$ of the cosmic string is much less than $L$.

It is natural to ask if gravitational back-reaction might prevent our experiment from taking place.  There are potential issues at both large and small scales, but it is easy to check that both are avoided.
Large-scale back-reaction is shown to be small by noting that the total (asymptotically-measured) energy of our apparatus is small compared to the mass $M_{BH}$ of the black hole.  Indeed, this energy satisfies

\begin{equation}
E_{\rm apparatus} \lesssim
 \mu r_0 + m_{\rm det}  = \epsilon^{-1} L^{-1} (r_s/\epsilon L +1) \sim \epsilon^{-2} r_s/L^2 \ll r_s/\ell_p^2.
\end{equation}
So since $d \ge 4$ we have
\begin{equation}
E_{\rm  apparatus} \ll   \frac{r_s}{\ell_p^2 } \left(\frac{r_s}{\ell_p} \right)^{d-4} \sim M_{BH},
\end{equation}
and there is no further restriction on our experiment\footnote{\label{lim} However, back-reaction does prevent one from placing an arbitrary number of such strings near the black hole.  This limits the number of mining processes that can run concurrently and thus the total rate at which energy can be extracted from the black hole.  Since there are $\sim (R/L)^{d-2}$  Hawking quanta at the scale $L$, one would like to use $N_A \sim (R/L)^{d-2}$ copies of our apparatus.  The constraint $E_{\rm apparatus} \ll M_{BH}$ then requires $L^d \gg r_s^2 \ell_p^{d-2}$ and
allows us to mine energy only at rates
$E/t \ll M_{BH}/t_{\rm extract}$ for $t_{\rm extract} = T_H^{-1} \left( {r_s}/{\ell_p} \right)^{2(1-2/d)}$  in agreement with \cite{FF}.  Similar arguments will appear in \cite{AB}. Without this constraint, one would obtain the Unruh-Wald result $t_{\rm extract} \sim T_H^{-1}$.}.

At small scales, one might ask whether our waiting detector is close enough to the black hole to be engulfed by even a small tide raised on the horizon by the gravitational field of our apparatus.  But since tidal effects are short-ranged ($\sim 1/r^{d-1}$), such a tide will be due mostly to the detector and the very bottom part of the string (within $\sim L$ of the horizon).  It can therefore be addressed using the Rindler approximation to the black hole geometry.  Dimensional analysis, the lack of any scales in Rindler space,  and the fact that the detector mass can enter only through $Gm_{\rm det}/T_H$ then imply that there can be no such effect for
\begin{equation}
\label{tide}
L \gg \ell_p (m_{\rm det}/T_{\rm loc})^{1/(d-2)} \sim \ell_p \epsilon^{-1/(d-2)} .
\end{equation}

Our discussion above involved the use of a cosmic string. For more mundane strings, one would be forced to consider whether the string is in fact strong enough to support its own weight and that of the detector being raised.    It turns out \cite{AB} that any sufficiently strong string acts much like a cosmic string.  But this observation raises a final concern:  As opposed to jump ropes and bicycle chains, the tension of a cosmic string is fixed once and for all.   We can choose parameters so that our detector is in static equilibrium at height $L$ (so that the upward pull from the string balances the gravitational attraction of the black hole), but this equilibrium is necessarily unstable.

Let us therefore suspend our detector on a pair of cosmic strings, instead of just one, so that the two strings meet at our detector with some non-zero angle.  The upward force then depends on the angle between the two strings.  As with a piano wire, the net force increases when the string is pulled downward.  This effect can be used to stabilize the detector at its operating location, and the detector can be raised and lowered by moving top ends of the strings along the Dyson sphere.

Moving the detector adiabatically slowly makes the process reversible so that no excess energy is left behind in the black hole.   In fact, one can perform the experiment well within the natural black hole evaporation time $t_{evap} \sim r_s^{d-1}/\ell_p^{d-2}$ of a $d \ge 4$ an asymptotically flat black hole without generating significant entropy.  This can be seen by first noting that \eqref{tide} implies, even if the detector were to fall through the horizon, that the formation of caustics is not relevant to the production of horizon entropy \cite{membrane,Amsel:2007mh}.  One may then use the Raychaudhuri equation parametrized by Killing time instead of affine parameter as in derivation of the physical process first law \cite{firstlaw} to write

\begin{equation}
\Delta A \sim r_s   \int dt \,dA\, \sigma^2,
\end{equation}
where we have taken the right-hand side of the Raychaudhuri equation to be dominated by the shear contribution
$\sigma^2$  generated by gravitational tides from the moving detector.   Compare with e.g. eqn. (2.7) of \cite{Amsel:2007mh}.  Since tidal effects decrease rapidly with distance, the integrals are dominated by the UV scale $L$ and we have
\begin{equation}
\frac{\Delta A}{\ell_p^{d-2}} \sim r_s    \frac{\ell_p^{d-2}m_{\rm det}^2}{L^{d-3}}   \left(\frac{r_s}{L} \frac{dL}{dt}\right)^2 \sim \frac{r_s^3}{\epsilon^2 Lt^2}   \left( \frac{\ell_p}{L}\right)^{d-2}   ,
\end{equation}
where $\frac{dL}{dt}$ represents a  typical value characterizing the motion of the detector at the scale $L$ (which is related to a typical velocity $v = \frac{r_s}{L} \frac{dL}{dt}$ seen by a typical freely-falling observer through the redshift factor $r_s/L$).  In the final step we have used $m_{det} = \epsilon^{-1} L^{-1}$ and we approximated $\frac{dL}{dt} \sim L/t$, where $t$ is the timescale of the experiment.    Since $L \gg \ell_p$, the result will be small whenever $t \gtrsim \sqrt{r_s^3/ \ell_p}$.

\end{document}